\documentclass{aa}
\usepackage{graphicx}
\usepackage{txfonts}
\usepackage{natbib}
\begin{document}
\voffset=1.0cm
\newcommand{\be}{\begin{equation}}
\newcommand{\ee}{\end{equation}}

\title{Importance of Compton scattering for radiation 
spectra of isolated neutron stars with weak magnetic fields}

\author{
V.~Suleimanov\inst{1,2},
K.~Werner\inst{1}}

\offprints{V.~Suleimanov}
\mail{e-mail: suleimanov@astro.uni-tuebingen.de}

\institute{
Institut f\"ur Astronomie und Astrophysik, Universit\"at T\"ubingen, Sand 1,
 72076 T\"ubingen, Germany
\and
Kazan State University, Kremlevskaja str., 18, Kazan 420008, Russia
}

\date{Received xxx / Accepted xxx}

   \authorrunning{Suleimanov \& Werner}
   \titlerunning{Compton scattering in isolated neutron stars}

\abstract
{}
{Emergent model spectra of neutron star atmospheres are widely used to
fit the observed soft X-ray spectra of different types of isolated neutron
stars. We investigate the effect of Compton scattering on the emergent spectra
of hot  ($T_{\rm eff} \ge 10^6$ K) isolated neutron stars with weak magnetic
fields.}
{In order to compute model atmospheres in hydrostatic and radiative
equilibrium we solve the radiation transfer equation with the Kompaneets
operator. We calculate a set
of models with effective temperatures in the range 1 - 5 $ \cdot 10^6$ K, with
two values of surface gravity ($\log~g$ = 13.9 and 14.3) and different
chemical compositions.}
{Radiation spectra computed with Compton scattering are softer than those
computed 
without Compton scattering at high energies ($E>$ 5 keV) for light elements
(H or He) model atmospheres. The Compton effect is more significant  in H
model atmospheres and models with low surface gravity. The emergent
spectra of the hottest ($T_{\rm eff} >3 \cdot 10^6$ K) model atmospheres can be
described 
by diluted blackbody spectra with hardness factors $\sim$ 1.6 - 1.9.
Compton scattering is less important in models with solar abundance of heavy
elements.}
{}


\maketitle
%

\section{Introduction}

Relatively young neutron stars (NSs) with ages $\le 10^6$ yr are 
hot enough ($T_{\rm eff} \sim 10^6$ K) and can be observed as soft X-ray
sources. Indeed, the thermal radiation of the isolated NSs
was discovered by the X-ray observatories {\it Einstein} and {\it EXOSAT}
\citep{Cheng.Helfand:83, Brinkmann.Ogelman:87, Cordovaetal:89}. At the present 
time, the thermal radiation of a few tens  of isolated NSs  of different
kinds, from anomalous 
X-ray pulsars to millisecond pulsars, are detected. The thermal spectra of
these objects can be described by blackbody spectra with (color) temperatures
from 40 to 700 eV (see, for example, \citealt{Mereghettietal:02}).

The nature of isolated NS surface layers is not known exactly. Under some
conditions (depending on surface temperature, magnetic field strength, and
chemical 
composition), a surface can be solid, liquid, or have a plasma envelope
\citep{Lai.Salpeter:97, Lai:01}. In the last case, the envelope
can be considered as an NS atmosphere, and 
the structure and emergent spectrum of this atmosphere can be computed by
using stellar model 
atmosphere methods \citep[e.g.\ ][]{Mihalas:78}. Such modeling has been performed by
many 
scientific groups, beginning with \cite{Romani:87},  for an isolated NS model
atmospheres without a magnetic field \citep{Zavlinetal:96, Rajagopal.Romani:96,
Werner:00, Gansickeetal:02, Ponsetal:02}, as well as for models with strong
($B > 10^{12}$ G) 
magnetic fields \citep{Shibanovetal:92, Rajagopaletal:97, Ozel:01, Ho.Lai:01,
Ho.Lai:03, Ho.Lai:04}. These
 model spectra were used to fit the 
observed isolated NS X-ray spectra (see review by \citealt{Pavlov.Zavlin:02}). 

One of the important results of these works is as follows. Emergent model spectra
of light elements (hydrogen and helium) NS atmospheres with low magnetic field
are significantly harder  than the
 blackbody spectra of the temperature $T_{\rm eff}$. These elements are
fully ionized in  atmospheres with $T_{\rm eff} \ge 10^6$ K. Therefore, the
true  opacity in these
atmospheres (mainly due to free-free transitions) decreases with photon energy
as $E^{-3}$. At high energies  electron scattering is larger than the true
opacity and photons emitted deep in the  atmosphere (where $T > T_{\rm eff}$)
escape after a few scatterings on electrons.  In the all of  previous
works, the model spectra of isolated NS were calculated with coherent 
 (Thomson) electron scattering taken into consideration. As a
result,  hard photons, which are emitted in the deep hot layers of the
atmosphere, escape without changing their energy. But if we take  Compton
scattering into
account, the hard photons will lose energy at each
scattering event. Therefore, such Compton down-scattering can affect  emergent
spectra of light elements model atmospheres of isolated NS.

It is well known that the Compton down-scattering determines the shape of
emergent model spectra of hotter NS atmospheres with $T_{\rm eff} \sim 2 \cdot
10^7$ K and close to the Eddington limit \citep{Londonetal:86, Lapidusetal:86,
Ebisuzaki:87, Zavlin.Shibanov:91, Pavlov.etal:91}. These model spectra
describe the 
observed X-ray spectra of X-ray bursting NS in low-mass X-ray binaries
(LMXBs),  and they are close to
diluted blackbody spectra with a hardness factor $f_c \sim$ 1.5 - 1.9
\citep{Londonetal:86, Lapidusetal:86, 
Ebisuzaki:87, Zavlin.Shibanov:91, Pavlov.etal:91}. 
But these model atmospheres with  Compton scattering taken into account are
not calculated for relatively cool atmospheres with $T_{\rm eff} < 10^7$
K. Therefore, at present, the effect of Compton scattering on the emergent
spectra of isolated NS
model atmospheres with $T_{\rm eff} < 5 \cdot 10^6$ K is not well known.   
  It is necessary to point out, that the diluted blackbody spectrum has
 the same spectral energy distribution as the black body spectrum with a given
 temperature $T_{\rm c}$, but it has a lower flux. As a result, the
bolometric flux of the diluted blackbody 
 spectrum is lower than the bolometric flux of the blackbody spectrum with
 temperature $T_{\rm c}$. The bolometric flux of the diluted blackbody
 spectrum corresponds to the effective temperature $T_{\rm eff} < T_{\rm
c}$. The hardness factor is determined as the ratio of these temperatures:
$f_{\rm c} = T_{\rm c}/T_{\rm eff}$.   

In this paper, we compute model atmospheres of NSs with  Compton
scattering taken into consideration and investigate the Compton effect on the
emergent model spectra of these atmospheres. We consider the importance of
Compton scattering qualitatively in \S\ref{s:compton}. 
  Our methods of calculation are outlined in
\S\ref{s:methods}, while results and conclusions are briefly discussed
in  \S\S\ref{s:results} and  \ref{s:conclusions}.

\section{Importance of Compton scattering}
\label{s:compton}

\begin{figure}
\includegraphics[width=1.0\columnwidth]{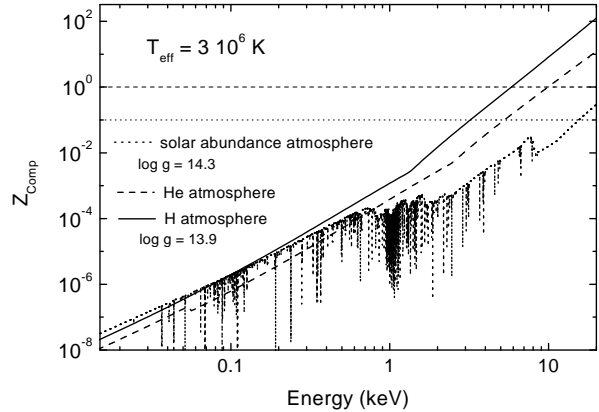}
\caption{\label{f:fig1} 
Comptonization parameter $Z_{\rm Comp}$ vs. photon energy for neutron star
model atmospheres with different chemical compositions.} 
\end{figure}

First of all, we consider the Compton scattering effect on emergent model
spectra of isolated NS atmospheres qualitatively. It is well known that in the
non-relativistic approximation ($h\nu, kT_{\rm e} << m_{\rm e} c^2$) the
relative photon energy lost due to a scattering event on a cool electron is 
\be
     \frac{\Delta E}{E} \approx \frac{h\nu} {m_{\rm e} c^2}.
\ee   
Each scattering event changes the relative photon energy by this value. It is
clear 
that the Compton scattering effect can be significant, if the final photon
energy change is comparable to the initial photon energy. Therefore, we
can define the Comptonization parameter $Z_{\rm Comp}$ (see also \citealt{Sul2006}):
\be Z_{\rm
Comp} = \frac{h\nu}{m_{\rm e}c^2} \max((\tau_{\rm e}^*)^2,\tau_{\rm
e}^*), 
\label{e:zcomp}
\ee 
where  $\max((\tau_{\rm
e}^*)^2,\tau_{\rm e}^*)$ is the number of scattering events that the photon
undergoes before escaping, and $\tau_{\rm e}^*$ is the Thomson optical depth,
corresponding to the depth where escaping photons of a given frequency
are emitted. 
We can expect that Compton effects on emergent spectra of NS model atmospheres
are significant if the Comptonization
parameter approaches unity \citep{Rybicki.Lightman:79}. Because of this we
compute  $Z_{\rm Comp}$ at different photon energies (see Fig.\,\ref{f:fig1}) for hot NS
model atmospheres with different chemical compositions. These models were
computed by using the method described in the next section, with Thomson electron
scattering.  It is seen from Fig.\,\ref{f:fig1} that the Comptonization parameter is
larger  (0.1 - 1) at high photon energies ($E>4 - 5$ keV) for H and He model
atmospheres. Therefore, we can expect a significant effect of Compton
scattering on the emergent spectra of these models. On the other
hand,  $Z_{\rm Comp}$ is small for the model  with solar chemical
composition of heavy elements. The Compton scattering effect on the emergent
spectrum  of this model has to be weak.

This qualitative analysis shows that Compton scattering can be significant for
light element model atmospheres of isolated NS and we investigated this
quantitatively in more detail. 

\section{Method of calculations}
\label{s:methods}

We computed model
atmospheres of hot, isolated NSs subject to the constraints 
of hydrostatic and radiative equilibrium assuming planar geometry
using standard methods \citep[e.g.\ ][]{Mihalas:78}.

The model atmosphere structure for an NS with effective temperature $T_{\rm
eff}$, surface gravity $g$, and given chemical composition is described by 
a set of differential equations. The first is the hydrostatic equilibrium
equation  
\be \label{e:hyd}
  \frac {d P_{\rm g}}{dm} = g -
4 \pi \int_0^{\infty}  H_{\nu} \, \frac{k_{\nu}+\sigma_\nu}{c} \, d\nu,
\ee
where 
\be
    g=\frac{GM_{\rm NS}}{R^2_{\rm NS}\sqrt{1-R_{\rm S}/R_{\rm NS}}},
\ee
and when $R_{\rm S}=2GM_{\rm NS}/c^2$ is the Schwarzschild radius of the NS,
$k_{\nu}$  opacity per unit mass due to free-free, bound-free, and 
bound-bound
transitions, $\sigma_{\rm e}$  the electron (Thomson) opacity, $H_{\nu}$ 
Eddington flux, $P_{\rm g}$  a gas pressure, and the column density $m$ is
determined as
\be
     dm = -\rho \, dz \, .
\ee
The variable $\rho$ denotes the gas density and $z$ is the vertical distance.

The second is the radiation transfer equation with
Compton scattering  taken into account  using the Kompaneets operator 
\citep{Kompaneets:57, Zavlin.Shibanov:91, Grebenev.Sunyaev:02}:
\begin{eqnarray} \label{rtr}
   \frac{\partial^2 f_{\nu} J_{\nu}}{\partial \tau_{\nu}^2} =
\frac{k_{\nu}}{k_{\nu}+\sigma_{\rm e}} \left(J_{\nu} - B_{\nu}\right) -
 \frac{\sigma_{\rm e}}{k_{\nu}+\sigma_{\rm e}} \frac{kT}{m_{\rm e} c^2}
\times \\ \nonumber
 x
\frac{\partial}{\partial x} \left(x \frac{\partial J_{\nu}}{\partial x} -
3J_{\nu} + \frac{T_{\rm eff}}{T} x J_{\nu} \left( 1 + \frac{CJ_{\nu}}{x^3}
\right) \right),
\end{eqnarray}
where $x=h \nu /kT_{\rm eff}$ is the dimensionless frequency,
$f_{\nu}(\tau_{\nu})$  the variable Eddington factor, $J_{\nu}$
 the mean intensity of radiation, $B_{\nu}$  the black body (Planck)
intensity,
$T$  the local
electron temperature,
 and $C=c^2 h^2~/~2(kT_{\rm eff})^3$. The optical
depth $\tau_{\nu}$ is defined as
\be
    d \tau_{\nu} = (k_{\nu}+\sigma_{\rm e}) \, dm.
\ee
These equations have to be completed both by the energy balance equation
\begin{eqnarray}  \label{econs}
 \int_0^{\infty} k_{\nu}\left(J_{\nu} - B_{\nu}\right) d\nu -
 \sigma_{\rm e} \frac{kT}{m_{\rm e} c^2} \times \\ \nonumber
\left( 4 \int_0^{\infty} J_{\nu} \,
d\nu - \frac{T_{\rm eff}}{T} \int_0^{\infty} x J_{\nu}
(1+\frac{CJ_{\nu}}{x^3}) \, d\nu \right)=0,
\end{eqnarray}
the ideal gas law
\be   \label{gstat}
    P_{\rm g} = N_{\rm tot} kT,
\ee
where $N_{\rm tot}$ is the number density of all particles, and also
by the particle and charge conservation equations.  We assume local
thermodynamical equilibrium (LTE) in our calculations, so the number
densities of all ionization and excitation states of all elements were
 calculated using Boltzmann and Saha equations. We take the
pressure ionization effects  into account on atomic populations using the occupation
probability 
formalism \citep{Hum.Mih:88} as  described by \citet{Lanz.Hub:94}.

For solving the above equations and computing the model atmosphere, we
used a version of the computer code ATLAS \citep{Kurucz:70,Kurucz:93},
modified to deal with high temperatures.  In particular, we take free-free and
bound-free transitions into
consideration  for all ions of the 15 most
abundant chemical elements, and  about 25\,000 spectral lines of these ions
are also included; see \citet{Ibragimov.etal:03}
and \citet{Swartz.etal:02} for further details.  This code was also
modified to account for Compton scattering (\citealt{Sul.Pout:06}; 
\citealt{Sul2006}).

The scheme of calculations is as follows.  First of all, the input
parameters of the model atmosphere  are defined: the effective temperature $T_{\rm
eff}$, surface gravity $g$, and the chemical composition.  Then a
starting model using a grey temperature distribution is calculated.
The calculations are performed with a set of 98 depth points $m_{\rm
i}$ distributed logarithmically in equal steps from $m\approx
10^{-7}$~g~cm$^{-2}$ to $m_{\rm max}$. The appropriate value of
$m_{\rm max}$ is found from the condition $\sqrt{\tau_{\nu,\rm
b-f,f-f}(m_{\rm max})\tau_{\nu}(m_{\rm max})} >$ 1 at all frequencies.
Here $\tau_{\nu,\rm b-f,f-f}$ is the optical depth computed with only the true
opacity (bound-free and free-free transitions, without scattering) taken into
consideration.  
Satisfying this equation is necessary for the inner boundary condition
of the radiation transfer.

 For the starting model, all number densities and opacities at all
depth points and all frequencies are calculated. We use 300 logarithmically
equidistant frequency points in our computations. The radiation transfer
equation (\ref{rtr}) is non-linear and is solved iteratively by the
Feautrier method \citep[][;~see also \citealt{Zavlin.Shibanov:91,
Pavlov.etal:91,Grebenev.Sunyaev:02}]{Mihalas:78}.  We use the last
term of Eq. (\ref{rtr}) in the form
$xJ_{\nu}^i(1+CJ_{\nu}^{i-1}/x^3)$, where $J_{\nu}^{i-1}$ is the mean
intensity from the previous iteration.  During the first iteration we
take $J_{\nu}^{i-1}=0$.  Between iterations we calculate the variable
Eddington factors $f_{\nu}$ and $h_{\nu}$, using the formal solution
of the radiation transfer equation in three angles at each frequency.
In the considered models  4-5 iterations are sufficient for achieving convergence.

We used the usual condition at the outer boundary
\be
    \frac{\partial J_{\nu}}{\partial \tau_{\nu}} = h_{\nu} J_{\nu},
\ee
where $h_{\nu}$ is the surface variable Eddington factor.
The inner boundary condition is
\be
   \frac{\partial J_{\nu}}{\partial \tau_{\nu}} =
 \frac{\partial B_{\nu}}{\partial \tau_{\nu}}.
\ee
The outer boundary condition is found from the lack of incoming
radiation at the NS surface, and the inner boundary condition is obtained
from the diffusion approximation $J_{\nu} \approx B_{\nu}$ and $H_{\nu}
\approx 1/3 \times \partial B_{\nu}/\partial \tau_{\nu}$.

The boundary conditions along the frequency axis are
\be  \label{lbc}
      J_{\nu} = B_{\nu}
\ee
at the lower frequency boundary ($\nu_{\rm min}=10^{14}$ Hz,
 $h\nu_{min}$ $\ll kT_{\rm eff}$), and
\be  \label{hbc}
x \frac{\partial J_{\nu}}{\partial x} - 3J_{\nu} + \frac{T_{\rm eff}}{T} x
J_{\nu} \left( 1 + \frac{CJ_{\nu}}{x^3} \right)=0
\ee
at the upper frequency boundary ($\nu_{\rm max}\approx  
10^{19}$ Hz, $h\nu_{\rm max} \gg kT_{\rm eff}$).  Condition
(\ref{lbc}) means that at the lowest energies the true opacity
dominates over scattering $k_{\nu} \gg \sigma_{\rm e}$, and therefore
$J_{\nu} \approx B_{\nu}$. Condition (\ref{hbc}) means that there is
no photon flux along the frequency axis at the highest energy.

The solution of the radiative transfer equation (\ref{rtr}) was checked
for the energy balance equation (\ref{econs}), together with the surface
flux condition
\be
    4 \pi \int_0^{\infty} H_{\nu} (m=0) d\nu = \sigma T_{\rm eff}^4 = 4 \pi
   H_0.
\ee
The relative flux error 
\be
     \varepsilon_{H}(m) = 1 - \frac{H_0}{\int_0^{\infty} H_{\nu} (m) d\nu},
\ee
and the energy balance error as  functions of depth
\begin{eqnarray}  \label{econs1}
 \varepsilon_{\Lambda}(m)=  \int_0^{\infty} k_{\nu}\left(J_{\nu} - B_{\nu}\right) d\nu -
 \sigma_{\rm e} \frac{kT}{m_{\rm e} c^2} \times \\ \nonumber
~~~~~~~~~~~~~~ \left( 4 \int_0^{\infty} J_{\nu} \,
d\nu - \frac{T_{\rm eff}}{T} \int_0^{\infty} x J_{\nu}
(1+\frac{CJ_{\nu}}{x^3}) \, d\nu \right)
\end{eqnarray}
were calculated, where $H_{\nu} (m)$ is radiation flux at any given depth $m$.
This quantity is found from
the first moment of the radiation transfer equation:
\be
    \frac{\partial f_{\nu} J_{\nu}}{\partial \tau_{\nu}} = H_{\nu}.
\ee
Temperature corrections were then evaluated using three different
procedures.  The first is the integral $\Lambda$-iteration method,
modified for Compton scattering, based on the energy balance equation
(\ref{econs}). In this method the temperature correction for a particular
depth is found from 
\be
     \Delta T_{\Lambda} = \frac{-\varepsilon_{\Lambda}(m)}{\int_0^{\infty}
\left[ (\Lambda_{\nu~diag}-1)/(1-\alpha_{\nu}\Lambda_{\nu~diag}) \right]
k_{\nu} (dB_{\nu}/dT)\,d \nu}.
\ee   
Here $\alpha_{\nu}=\sigma_{\rm e}/(k_{\nu}+\sigma_{\rm e})$, and
$\Lambda_{\nu~diag}$  is the diagonal matrix element of the $\Lambda$
operator  (see details in \citealt{Kurucz:70}).
This procedure is used in the upper atmospheric layers.  The
second procedure is the Avrett-Krook flux correction, which uses the
relative flux error $\varepsilon_{H}(m)$ and is performed in the deep layers.
And the third 
one is the surface correction, which is based on the emergent flux
error.  See \citet{Kurucz:70} for a detailed description of the 
methods.

The iteration procedure is repeated until the relative flux error is
smaller than 1\%, and the relative flux derivative error is smaller
than 0.01\%. As a result of these calculations, we obtain a
self-consistent isolated NS  model atmosphere, together with the emergent
spectrum of radiation.

Our method of calculation was tested on a model for X-ray bursting NS
 atmospheres \citep{Pavlov.etal:91, Madej.etal:04}. We found that our models are
in good agreement with these  calculations. 

\section {Results}
\label{s:results}

\begin{figure}
\includegraphics[width=1.0\columnwidth]{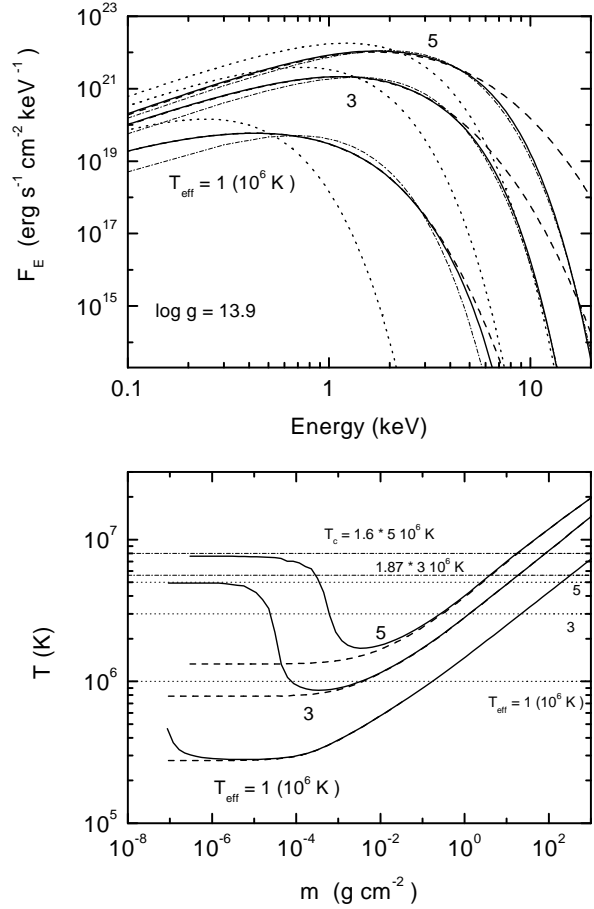}
\caption{\label{f:fig2} 
{\it Top panel:} Emergent (unredshifted) spectra of  pure hydrogen, low gravity
 ($\log g$=13.9) NS 
model atmospheres. Solid curves - with Compton effect; dashed curves - without
Compton effect; dotted curves - blackbody spectra; thin dash-dotted curves -
diluted blackbody spectra with hardness factors 2.79, 1.87, and 1.6 for models
with $T_{\rm eff}$ = 1,3 and 5$\cdot 10^6$ K. {\it Bottom panel:} Temperature structures
of the corresponding model atmospheres. Effective and color temperatures are
shown by dotted and dash-dotted lines, respectively.
}
\end{figure}

\begin{figure}
\includegraphics[width=1.0\columnwidth]{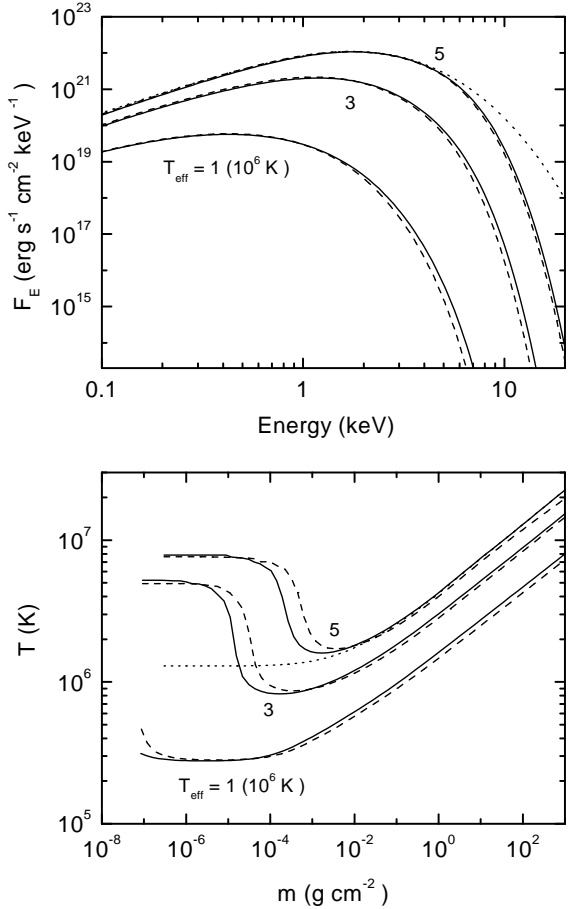}
\caption{\label{f:fig3}
{\it Top panel:} Emergent (unredshifted) spectra of pure hydrogen  NS model
atmospheres with different surface gravities (solid curves - high gravity 
($\log g$ = 14.3)
models; dashed curves - low gravity models ($\log g$ = 13.9)). For
comparison the model spectra 
without Compton effect are shown for the hottest high gravity model (dotted
curve). {\it Bottom
panel:} Temperature structures of the corresponding model atmospheres.
}
\end{figure}

\begin{figure}
\includegraphics[width=1.0\columnwidth]{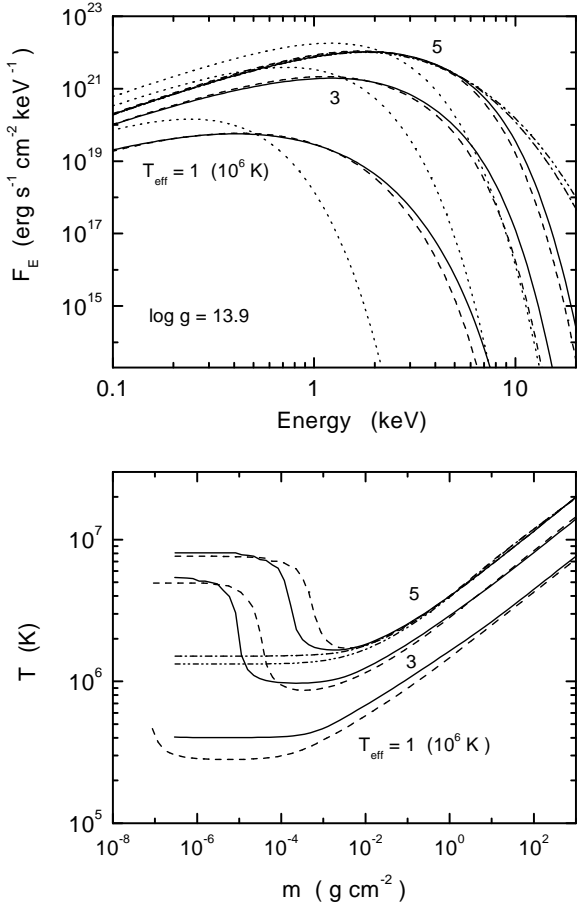}
\caption{\label{f:fig4} 
{\it Top panel:} Emergent (unredshifted) model spectra of pure helium, low
gravity  ($\log g$ = 13.9) NS 
model atmospheres. For comparison the model spectra of pure hydrogen
atmospheres are 
shown by dashed curves. Model spectra of the hottest atmospheres without
Compton effect are shown by dash-dotted and dash-dot-dotted curves. {\it
Bottom panel:} Temperature structures of the corresponding model atmospheres.
} 
\end{figure}

We use this method to calculate  a  set of hydrogen and helium isolated NS model atmospheres
with 
effective temperatures 1, 2, 3, 5 $\cdot 10^6$ K and surface gravities
$\log~g$ = 
13.9 and 14.3 was calculated. Models with Compton scattering and
Thomson scattering  were computed for comparison. Part of the results
 are presented in Figs.\,\ref{f:fig2} - \ref{f:fig4}.

 The Compton effect is significant for the
 spectra of hot ($T_{\rm eff} \ge 3 \cdot 10^6$ K) hydrogen model
atmospheres at high energies (Fig.\,\ref{f:fig2}). The hard emergent photons lose energy
and heat the upper layers of the atmospheres due to interactions with
electrons. As a result, the high-energy tails of the emergent spectra become
similar to Wien spectra. The temperature also increases in the upper layers of
 the model atmospheres and chromosphere-like structures appear. Temperatures
of these structures are close  to the color temperatures of the Wien tails of the
 emergent spectra.
 Moreover, the overall  emergent model spectra of high temperature atmospheres in a first
 approximation can be presented as diluted blackbody spectra with color
 temperatures that are close to Wien tail color temperatures
\be
     F_{\rm E} = \frac{\pi}{f_{\rm c}^4} B_{\rm E} (T_{\rm c}), ~~~~~~ T_{\rm
c}= f_{\rm c} T_{\rm eff}, 
\ee   
where $f_{\rm c}$ is hardness factor  and $F_{\rm E}$ is connected to
 $H_{\nu}$ as follows:
$$
        F_{\rm E} = 4 \pi H_{\nu}~ \frac{d\nu}{dE},~~~~~
        \frac{d\nu}{dE} = \frac{1.6022\cdot 10^{-9} {\rm erg}/{\rm keV}}{h}. 
$$

 These results are similar to those obtained
for model atmospheres and emergent spectra of X-ray bursting NS in LMXBs
\citep{Londonetal:86, Lapidusetal:86,
Ebisuzaki:87, Madej:91, Pavlov.etal:91, Madej.etal:04}. 

\begin{table*}
\begin{minipage}{180mm}
\begin{center}
\caption{The hardness factors $f_c$ for hydrogen and helium NS model
atmospheres with different $\log g$ computed for models with Compton (C) and
Thomson (T) scattering. The values in the brackets show the
deviations $\varepsilon$ of the model spectra from diluted blackbody spectra
with color 
temperature $T_{\rm c} = f_c T_{\rm eff}$ (see Eq. \ref{dv}).}
\label{tab1}
\begin{tabular}{lcccccccc}
 \hline
 \hline 
 & \multicolumn{4}{c}{Hydrogen models} & \multicolumn{4}{c}{Helium models} \\ 
\hline 
 $T_{\rm eff}$ & \multicolumn{2}{c}{$\log~ g$ = 13.9} &
 \multicolumn{2}{c}{$\log~ g$ = 14.3} & \multicolumn{2}{c}{$\log~ g$ = 13.9}&
 \multicolumn{2}{c}{$\log~ g$ = 14.3}\\
 & C & T & C & T & C & T & C & T \\
 \hline
 $1 \cdot 10^6$ K & 2.79 & 2.69 & 2.86 & 2.81 & 2.83 & 2.81 & 2.83 & 2.82 \\
    & (0.30) & (0.32) & (0.32) & (0.33) & (0.35) & (0.35) & (0.37) & (0.36)\\
 $2 \cdot 10^6$ K & 2.20 & 2.16 & 2.37 & 2.28 & ... & ... & ... & ... \\
    & (0.10) & (0.20) & (0.14) & (0.20) &  &  &  & \\
 $3 \cdot 10^6$ K & 1.87 & 1.91 & 1.97 & 1.99 & 2.05 & 2.04 & 2.15 & 2.12 \\
    & (0.05) & (0.15) & (0.07) & (0.14) & (0.09) & (0.13) & (0.11) & (0.14)\\
 $5 \cdot 10^6$ K & 1.60 & 1.64 & 1.64 & 1.67 & 1.67 & 1.69 & 1.72 & 1.73 \\
    & (0.03) & (0.08) & (0.03) & (0.07) & (0.04) & (0.06) & (0.04) & (0.06)\\

\hline
\hline \end{tabular} \end{center} 
\end{minipage}

\end{table*}

The Compton scattering effect on the emergent model spectra  of high gravity
atmospheres is less significant (Fig.\,\ref{f:fig3}). The reason is a relatively small
contribution of electron scattering to the total opacity in high gravity
atmospheres compared to low gravity ones. The mass density in the high
gravity 
models is higher, and the opacity coefficient (in cm$^2$/g) is independent of
the density for electron scattering and proportional to the density for
free-free transitions. In the presented models, H and He are practically
fully ionized. Therefore, free-free transitions dominate the true
opacity.

The Compton scattering effect on helium model atmospheres is also less
significant than on hydrogen model atmospheres with the same $T_{\rm eff}$ and
$\log~g$ (Fig.\,\ref{f:fig4}). The reason is the same as in the case of high gravity
models. The ratio of electron scattering to a true opacity is smaller in the
helium models. It is interesting to notice that, in the case of
Thomson scattering, the emergent model spectra of helium atmospheres are softer
than those of the hydrogen atmospheres. But in models where 
Compton scattering is taken into consideration, the spectra of helium atmospheres
are harder than the spectra of hydrogen models.

We fitted the calculated emergent spectra of this set of models by the diluted blackbody
spectra in energy range 0.2 - 10 keV. We computed the value
\be \label{dv}
    \varepsilon = \frac{1}{N}~ \Sigma_{j=1}^N ~\left(1-\frac{B_{\rm E,\,j}(f_c
T_{\rm eff})}{F_{\rm E,\,j}}\right)^2
\ee
as a measure for the deviation of the model spectrum $F_{\rm E}$  from the
diluted 
blackbody spectrum $B_{\rm E}$. Here $N$ is the number of energy points  E,\,j
between 0.2 and 10 keV.
Corresponding hardness factors  are
presented in the Table 1 together with $\varepsilon$. Spectra of the hot
models with Compton  scattering
are better approximated by a diluted black body, especially for hydrogen
models. But the hardness factors are changing only slightly. 
Data from Table 1 can be used for interpretating of observed
isolated NS X-ray spectra. The hardness factor can be found from these data if
the observed color temperature $T_c^{\infty}$ and gravitational redshift $z$
of the NS are known.
As soon as $f_{\rm c}$ is determined, the apparent
NS radius $R_{\infty}$ can be found from the observed flux $f_{\rm E}$ with
the relation:
\be
      f_{\rm E} = \frac{\pi}{f_{\rm c}^4} B_{\rm E} (T_{\rm c}^{\infty}) \frac{R_{\infty}^2}{d^2},
\ee
where $d$ is distance to the NS.

\begin{figure}
\includegraphics[width=1.0\columnwidth]{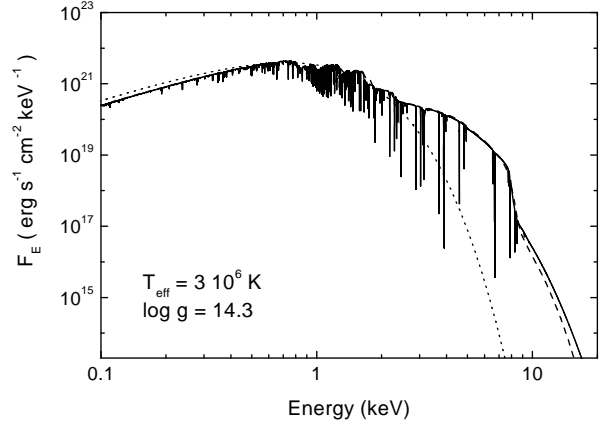}
\caption{\label{f:fig5} 
Emergent (unredshifted) model spectra of high gravity  ($\log g$ = 14.3)
NS atmospheres with the
solar abundance of 15 most abundant heavy elements with (dashed curve) and
without (solid curve) Compton scattering. The dotted curve is the corresponding
blackbody spectrum.
} 
\end{figure}

\begin{figure}
\includegraphics[width=1.0\columnwidth]{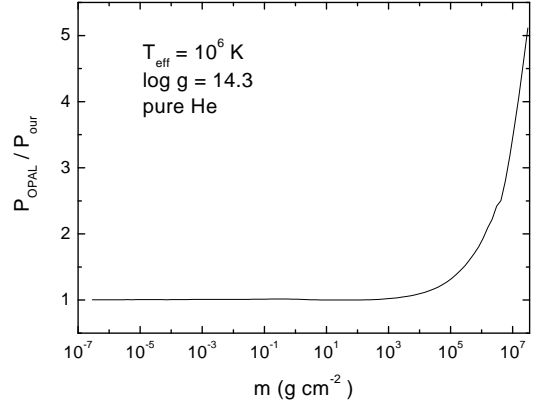}
\caption{\label{f:fig6} 
 Ratio of the gas pressure obtained by  interpolation from OPAL EOS tables
(see text) to the gas pressure from our calculations for the pure He model with
$T_{\rm eff}$ = 10$^6$ K and $\log g$ = 14.3. 
} 
\end{figure}

We also computed one isolated NS model atmosphere with solar chemical abundance and 
$T_{\rm eff} = 3 \cdot 10^6$ K and $\log~g$=14.3 (see Fig.\,\ref{f:fig5}). The model was
calculated with Thomson and Compton scattering and we found that the Compton
effect on the emergent spectrum is very small.

 In our calculations we ignored deviations from the ideal gas EOS
(\ref{gstat}) other than the ionization by pressure. We checked this 
approximation as follows. We found gas pressures
from the temperatures and densities at all depth points for a
pure helium atmosphere with $T_{\rm eff}$ = 10$^6$ K and $\log g$ = 14.3 
 by interpolation from the OPAL EOS
tables\footnote{http://www-phys.llnl.gov/Research/OPAL/}. We compared the obtained gas
pressures with the values calculated in our model. The ratio of these
two gas pressures are shown  in Fig.\,\ref{f:fig6}. We find that the OPAL EOS gas
pressure is equal  to the gas pressure in our model with an accuracy of about 1 \% up to
column density $\sim$ 10$^3$ g cm$^{-2}$. At larger depths the OPAL
EOS gas pressure is significantly larger due to partial electron degeneracy. All of the
emergent photons are emitted at column densities lower than 10$^{3}$ g
cm$^{-2}$ (for example, photons with energy 10 keV are emitted at column
density $\sim$ 20 g cm$^{-2}$).  The distinction between the OPAL EOS and
(\ref{gstat}) must be 
smaller for models with larger $T_{\rm eff}$ and for pure hydrogen models.    
Therefore, the EOS deviation  from
(\ref{gstat}) at the deepest atmospheric layers cannot change the emergent model
spectra that we calculated here.

\section{Conclusions}
\label{s:conclusions}

We have presented the results of calculations for the hot model atmospheres of isolated NSs
with 
low magnetic fields and different chemical compositions. The Compton effect is
taken into account. We investigated the  importance of Compton scattering for
the emergent spectra of these models. The main conclusions follow.
 
Emergent model spectra of hydrogen and helium NS atmospheres
with {\bf $T_{\rm eff} \ge 1 \cdot 10^6$} K are changed by the Compton effect at high energies
($E > 5$ keV), and spectra of the hottest ($T_{\rm eff} \ge 3 \cdot 10^6$ K)
model atmospheres  can be described by diluted blackbody spectra with
hardness factors $\sim$ 1.6 - 1.9. At the same time, however, the spectral energy
distribution  (SED) of these models are not significantly changed at the maximum of
the 
SED (at energies 1-3 keV), and effects  on the color temperatures are
not strong.

The Compton effect is the most significant for hydrogen
model atmospheres and in low gravity models.  Emergent model
spectra of NS atmospheres with solar metal abundances  are affected by
Compton effects only very slightly.

\begin{acknowledgements}

VS thanks DFG for financial support (grant We 1312/35-1) and the Russian FBR
(grant 05-02-17744) for partial support of this investigation.

\end{acknowledgements}


\bibliographystyle{aa}
\bibliography{6174text}

\end{document}